# A 3-Dimension Model of Volcanism Volatiles Contribution on Atmospheric Chemical Abundance of Habitable Planets

Zihan Huang*[1], Xing Zhou Q.[2]

1. Department of Physics and Astronomy, University of Manchester, Manchester M13 9PL, UK
2. Department of Physics, University of Chinese Academy of Science, Beijing 100049, CN

**Abstract**：

The volcanism plays an important part in mass exchange circle to bring matter from core of planet to atmosphere. Thus, it is a possible method to research the change of elements abundance in atmosphere by modeling the process of volatiles from volcanism get through and mix in atmosphere, which is the focused point of this article. This article penetrates from the generation of volatiles, talks the species, mass, and mole fractions of different typical elements in magma. Then a diffusion progress model of volatiles was built to quantify the abundance of elements with altitude. And the quantitative models of element abundance at different heights are obtained.

**Key words**：Chemical abundance; Atmosphere; Climate model; Volcanism; Astrochemistry

Volcanic activity is likely to occur on massive planets with plate tectonics, as they have the conditions to melt silicate mantle which could generate magma [1]. There are volatiles generated during the magma eruption, as the magma contains rich nonmetallic matter [2], which could be oxidized when contact with air or oxides and vapored by high temperature and formation pressure change [3]. Some elements in volatiles can stable exist in atmosphere after photochemical action with radiation or chemical reaction with composition in atmosphere [4], and then cause a notable abundant change during long term climate evolution. While causing changes in the abundance of atmospheric elements, the composition of magma also changes with different stages of the evolution of the interior of the planet. Volcanic activity acts as a channel to communicate the core of the planet and the atmosphere, which leads to indirect knowledge of the evolution process of the interior of the planet through spectral and other atmospheric observation methods. Therefore, to understand this channel, it is necessary to clear the impact of volcanic activity on the atmosphere, which can be specific to the contribution of volcanism to the abundance of elements in the atmosphere, as the abundance can be measured by spectrum which is commonly used [5]. This article will discuss and build the existence model of volatile matter in the atmosphere from the volcanic eruption process, the formation of volatile matter, and the diffusion of volatile matter.

# 1 Background

## 1.1 Generation of Volatiles

The common nonmetallic elements in magma are phosphorus (P), sulfur (S), silicon (Si), nitrogen (N), carbon (C), oxygen (O), hydrogen(H), and halogen group elements [6]. Among them, P, S, N, C, O, H and chlorine (Cl) are the elements that are easy to form gaseous molecules in the earth's atmospheric environment. These elements are also considered in this article when studying Earth like planets. In the atmospheric environment of Earth, the volatile substances that can be formed between every two elements are listed in Table.1. There are also volatile molecules formed by over three elements like carbonyl sulfide (COS) [6], however, these are not be considered in this article to simplify the model. Free radical molecules, such as cyapho radical (CP), which can exist at high altitude and are generated by photochemical action, are also not considered here, as they are not stable enough to exist at crater with high temperature.

Table.1 Two-element volatile molecules

|    | P | S | N | C | O | H | Cl |
|----|---|---|---|---|---|---|----|
| P  |   | $PS_x$ | $PN_x$ |   | $PO_x$ | $PH_3$ | $PCl_x$ |
| S  | - | S |   | $CS_2$ | $SO_x$ | $H_xS$ | $SCl_2$ |
| N  | - | - | $N_2$ | $(CN)_2$ | $NO_x$ | $H_xN$ | $NCl_3$ |
| C  | - | - | - |   | $CO_x$ | $CH_x$ | $CCl_4$ |
| O  | - | - | - | - | $O_x$ | $H_2O$ | $ClO_x$ |
| H  | - | - | - | - | - | $H_2$ | $HCl$ |
| Cl | - | - | - | - | - | - | $Cl_2$ |

Although chemical equilibrium models show that N-2, CO, S-2, CS2, S2Cl, Cl, Cl-2, and COS could be among the most abundant volcanic gases on some planets [6], the amount of all these molecules in Table.1 generated will be evaluated by the Gibbs energy function of their formation from single substances, which are from decomposition of mineral. Therefore, the amount of volatile matter produced will be related to the abundance of elements in the core of planet.

**1.2 Eruption of Volatiles**

Before talking about the diffusion of volatiles, the source of eruption needs to be defined. It is inappropriate to consider the source as a point which erupt only vertically. In the process of volcanic eruption, the crater area and eruption pressure will lead to the change of the ejection angle, which will affect the speed, area and height of the ejecta. It is not able to define an area of volatiles eruption directly, as gas are not able to see. However, there are small pyroclasts (less than 2 cm) in the uprising gas with terminal velocities which are ignorable when be compared with the gas velocity, meaning that the ejection velocity of clasts can be seen as equal to the velocity of the gas [2][7][8]. Then the range of pyroclasts spread can be considered as the area volatiles reached. The range of the pyroclasts spread can be obtained from satellite photos, which is an obvious shape in Fig.1(B) as an example, or calculated from ballistic trajectory of pyroclasts. Then, the volatiles can be seen as spreading evenly to all around from the edge of the hemisphere whose radius is the range of the pyroclasts spread.

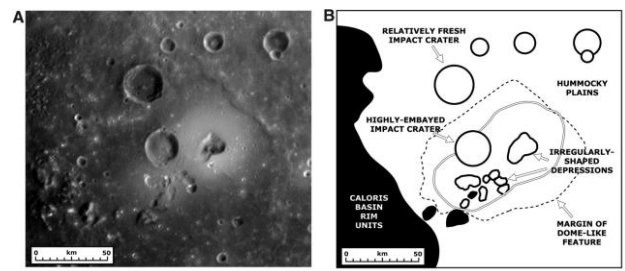

Fig.1(A) Surface of deposition area near volcano on Mercury, photographed by MESSENGER; (B)Sketch shape of the area [9].

The velocity of volatiles at the edge of the hemisphere can be denoted by the speed of volatiles generated at the vent of volcano. The velocity can be seen as the initial velocity of the volatiles diffusion.

**1.3 Diffusion of Volatiles**

When it turns to motion of volatiles in atmosphere, a reference frame need to be chosen. To simplify the model, this article will build the model by separating horizontal and vertical diffusion. In the model of vertical diffusion, the temperature and pressure function of height is the key factor need to focus on. In the horizontal model, the temperature and pressure gradient of latitude and longitude is the main factor that affect diffusion. As the thermal diffusion model will be used mainly in this article, but not the diffusion driven by concentration gradient [10]. The diffusion will be transfer to a rotating reference frame as the place considered at here is a planet, which is a sphere.

There are also other effects caused by rotating of planets need to be dealt, such like the geostrophic approximation when considering the area far from the surface of planet.

**1.4 Overview of Models**

As it shown in fig.2, this article simplified the process of volatile matter entering the atmosphere to several processes the volatiles need to go through, which were fitted to different models from inside to outside, they are source area (vent), expansion, initialization, deceleration, and diffusion.

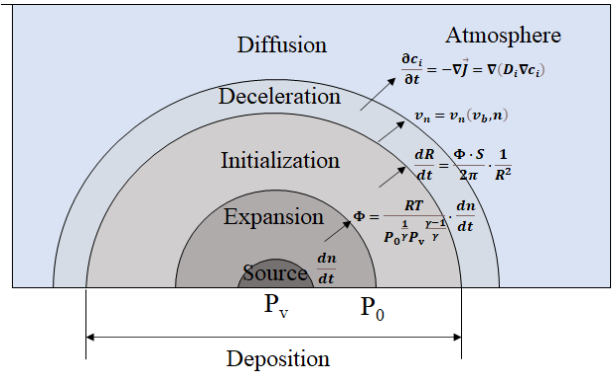

Fig.2 The models overview of this article.

## 2  Model Build

**2.1 Chemical Composition**

For each kind molecules listed in Table.1, the amount of generation during a term can be summed by all possible chemical components, here consider $PS_x$ as an example, which is denoted as formulation (F.1) given by models built in previous article [11].

$$n = \sum n_{PS_x} \qquad (F.1)$$

where $n_{PS_x}$ is the $x^{th}$ kind of possible chemical components of $PS_x$.

Thus, as the model provides the amount and ratio of different molecules at the source, where they were considered uniformly in the hemisphere around the vent of volcano, the following is to build the model about the diffusion of the molecules.

**2.2 Initial Velocity of Volatiles**

It is much easier to calculate the velocity of volatiles when there are satellites photos, those photos can provide direct radius scale, and then there are relationship in formulation (F.2) between the velocity of volatiles at the edge and the radius [2].

$$R = \frac{v^2}{g}\sin 2\theta \qquad (F.2)$$

where $R$ (m) is the maximum distance the volatiles can travel, which driven by eruption not diffusion, $v$ (m s$^{-1}$) is the velocity of volatiles at the vent, $g$ (m s$^{-2}$) is the gravitational acceleration at the surface of the planet, $\theta$ (degree) is the ejection angle of the volatiles when erupting from the vent.

When there are no direct evidences on the radius of eruption, a following simple model can be built to evaluate the velocity of volatiles at the edge. Consider there is a hemisphere with radius $R$ (m). The $dR$ is the differential of radius changed during dt. The $dV$ is the volume changed during $dt$, which is the dark area shown in the Fig.3.

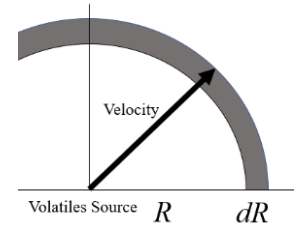

Fig.3 The relation between velocity at edge and radius

As the volume change was caused by volatiles which produced from source, assume the rate of volume of volatiles produced is $\Phi$ (m³ yr$^{-1}$ m$^{-2}$), there are relationship in (F.3) can be built based on Fig.3.

$$dV = \frac{2}{3}\pi((R+dR)^3 - R^3) = \Phi \cdot S \cdot dt \qquad (F.3)$$

where $S$ is the area (m²) of vent of volcano. When considering a long term, the $\Phi$ is a function of $t$, the formulation (F.3) can be denoted as (F.4).

$$dV = S \int \Phi \, dt \tag{F.4}$$

The formulation (F.3) is a first-order nonlinear inhomogeneous differential equations, which is normally hard to provide analytical solutions. However, if the expansion of volatiles is without residual, the velocity of every volatile at the edge can be considered from the process of the volatile traveling from the source to the edge. As all the volume generated during dt was considered as the volume of the hemisphere, there is formulation (F.5).

$$\frac{2}{3}\pi R^3 = \Phi \cdot S \cdot t \tag{F.5}$$

The velocity of the edge expansion of the hemisphere can be denoted by formulation (F.6).

$$\frac{dR}{dt} = \left(\frac{\Phi \cdot S}{18\pi}\right)^{\frac{1}{3}} t^{-\frac{2}{3}} \tag{F.6}$$

It can be simplified to formulation (F.7) by applying the relationship between $R$ and $t$.

$$v_0 = \frac{dR}{dt} = \frac{\Phi \cdot S}{2\pi} \cdot \frac{1}{R^2} \tag{F.7}$$

which $v_0 = \frac{dR}{dt}$ is the velocity of volatiles at the edge, a distance of $R$ from the center point of the vent, it was considered as the initial velocity of volatiles diffusion in this article. Thus, the velocity of volatiles at the vent, which relates to complicate parameters like ejection angle, was skipped through the above.

**2.3 Rate of Volatiles Generated**

To support the above model, the rate of volume of volatiles generated $\Phi$ in the formulation (F.3) need to be connected with amount of chemical composition $n$ in formulation (F.1). The ideal gas equation of state (F.8) is to denote the relation between the amount of substance and volume.

$$PV = nRT \tag{F.8}$$

where $P$ (MPa) is the pressure, $V$ (m³) is the volume, $n$ (mol) is the number of moles of gas, $R$ (8.314 kJ/K) is the universal gas constant, $T$ is the temperature. Thus, by using formulation (F.8) at the vent, when there are $n$ mole gas generated per second, there are $V$ volume of volatiles produced per second. The $P$ and $T$ are both the parameters at the vent of volcano. However, the rate of volume here can still not be considered as the rate of volume generated from the source, as there is pressure gradient which from vent to the edge of hemisphere. The volume change during this process can be approximated to an adiabatic expansion. There is equation (F.9) to represent the adiabatic process of ideal gas.

$$PV^\gamma = constant \tag{F.9}$$

where $\gamma$ is the adiabatic index ratio defined as formulation (F.10).

$$\gamma = \frac{f + 2}{f} \tag{F.10}$$

where $f$ is the degrees of freedom, which are 3 for monatomic molecules, 5 for linear molecules, 6 for non-linear molecules [12]. The vibration degrees freedom does not need to be considered under a temperature of the expansion process talked here, as it is well below the temperature to provide the energy to activate the vibrational modes [13]. Thus, the volume rate of volatiles generated $\Phi$ at the edge after the expansion can be expressed as formulation (F.11).

$$\Phi = \frac{RT}{P_0^{\frac{1}{\gamma}} P_v^{\frac{\gamma-1}{\gamma}}} \cdot \frac{dn}{dt} \tag{F.11}$$

where $P_0$ is the pressure at the edge of expansion, which was considered as sea level air pressure, $P_v$ is the pressure at the vent of volcano, $\frac{dn}{dt}$ is the rate of moles of volatiles generated at the vent.

**2.4 Vertical Thermal Diffusion**

To define the rate of thermal diffusion in gas, there is a diffusion model in gas, the first Fick's law, as formulation (F.12) [14].

$$\vec{J} = -lv\frac{\vec{\nabla}n}{3} = -D\vec{\nabla}n \qquad (F.12)$$

$J$ (mol m$^{-2}$ s$^{-1}$) depends on the speeds of molecule and the gradient in the concentration of the diffusing species. $D = lv/3$ is called the diffusion coefficient, which is relevant to temperature and pressure. When considering the concentration of gas varies with time, there is formulation (F.13), the second Fick's law [15].

$$\frac{\partial c_i}{\partial t} = -\nabla \vec{J} = \nabla(D_i \nabla c_i) \qquad (F.13)$$

where $c_i$ is the concentration of component $i$ at the considered location. $D_i$ is the diffusion coefficient of component $i$ in air of the planet, which is a function of temperature (F.14) [16][17].

$$D_i = \frac{AT^{\frac{3}{2}}}{P\sigma_{i-air}^2 \Omega}\sqrt{\frac{1}{M_i} + \frac{1}{M_{air}}} \qquad (F.14)$$

where $A$ (1.859×10$^{-3}$) is an empirical coefficient, $T$ is the temperature (K), $P$ is the pressure (atm), $\sigma_{i-air} = 0.5(\sigma_i - \sigma_{air})$ is the average collision diameter (Å) [18], $\Omega$ is a temperature-dependent collision integral, $M$ is the molar mass (g mol$^{-1}$) [18].

As $T$ and $P$ are functions of vertical distance $z$, the formulation (F.13) can be written as (F.15) when it only considers the vertical diffusion:

$$\frac{\partial c_i}{\partial t} = \frac{\partial}{\partial z}\left(D_i \frac{\partial c_i}{\partial z}\right)$$
$$= \left(\frac{\partial D_i}{\partial T} \cdot \frac{\partial T}{\partial z} + \frac{\partial D_i}{\partial P} \cdot \frac{\partial P}{\partial z}\right) \cdot \frac{\partial c_i}{\partial z} + D_i \frac{\partial^2 c_i}{\partial z^2}$$
$$(F.15)$$

The formulation (F.15) is the differential equation of molecular mass transfer of vertical diffusion.

**2.5 Horizontal Diffusion**

Whether it is vertical or horizontal diffusion, the driving force is always the reduce of freedom energy [19]. The horizontal diffusion model can apply the same equation of Fick's law (F.13). The temperature was considered as a constant during the horizontal diffusion without convection. Gradient of pressure was mainly considered. Thus, there is following formulation (F.16) to express the differential equation of molecular mass transfer of horizontal diffusion:

$$\frac{\partial c_i}{\partial t} = \nabla(D_i \nabla c_i)$$
$$= \frac{\partial D_i}{\partial P} \cdot \nabla c_i \cdot \nabla P + D_i \cdot \nabla^2 c_i \qquad (F.16)$$

**2.6 Deceleration Zone**

The diffusion above was considered without the initial velocity of volatiles, so it is necessary to define a deceleration zone between expansion and diffusion. It was considered that the deceleration was caused by collision between molecules here. Consider the velocity of volatile molecule finally reduce to the root-mean-squared velocity $v_s$ at a distance as $s$ (m). The $s$ is the radius of the deceleration zone defined in this section. The amount of atoms collided by a single volatile molecule during $\Delta s$ that is $N_{\Delta s}$, which can be denoted as formulation (F.17).

$$N_{\Delta s} = \frac{N_A}{M_{air}} \int_0^{\Delta s} \sigma_{i-air} \rho(s) ds \qquad (F.17)$$

where $N_A$ (6.022×10$^{23}$ mol$^{-1}$) is Avogadro constant, $\rho(\Delta s)$ (g m$^{-3}$) is the density of air at $\Delta s$, which is a function of temperature and pressure, it can be denoted as (F.18) by (F.8). Thus, $N_{\Delta s}$ can be written as $N_{\Delta s} = N_{\Delta s}(\Delta s)$.

$$\rho(\Delta s) = \frac{P(\Delta s)\overline{M_{air}}}{RT(\Delta s)} \qquad (F.18)$$

To a unidimensional completely elastic collision deceleration process, the velocity of volatiles reduces from initial velocity to root-mean-squared velocity at the terminal point, there are following equations (F.19) during each collision.

$$mv_b = mv_n + \bar{m}v_*$$
$$mv_b^2 + 6kT_n = mv_n^2 + \bar{m}v_*^2 \qquad (F.19)$$

where $m$ (g) is the mass of a single volatile molecule, $v_b$ (m s$^{-1}$) is the velocity of volatile molecule before n$^{th}$ collision, $v_n$ is the velocity of volatile molecule after n$^{th}$ collision, $\bar{m}$ is the average mass of a single air molecule, $v_*$ is the velocity of air molecule after n$^{th}$ collision. $T_n$ (K) is the temperature of the place n$^{th}$ collision happen, which is a function of $\Delta s$. The vibrational velocities of volatiles and air molecules were considered as statistic average quantities, which are non-direction vectors with modules as root-mean-squared velocities. So the average momentum of air molecule before collision is zero, and the translational kinetic energy is $1.5kT$ for both volatiles and air molecules. $k$ (1.381×10$^{-23}$ J/K) is Boltzmann constant. The relation between $T$ and $v_1$ can be derived by (F.19), which is formulation (F.20).

$$mv_b^2 + 6kT_n = mv_n^2 + \frac{(mv_b - mv_n)^2}{\bar{m}} \quad (F.20)$$

By ordering $N_{\Delta s}$ equal to n, the distance $\Delta s_n$ of the n$^{th}$ collision was solved by (F.17), then the temperature $T_n$ is solved by the function of $T$ on $s$, which was expressed as $T_n = T(\Delta s_n) = T(n)$. Thus, $v_n$ can be written as formulation (F.21), which is from equation (F.20).

$$v_n = v_n(v_b, n) \quad (F.21)$$

(F.21) is a general recursive function. To the first collision, the initial condition is $v_b = v_0$, n=1. $v_0$ is the initial velocity calculated by (F.7). To the second collision, it is done by applying $v_b = v_{n=1}$, n=2. To the last collision, there are following equations (F.22) as final conditions.

$$\begin{cases} v_n = v_s = \sqrt{\dfrac{3kT_s}{m}} \\ n = N_s = \dfrac{N_A}{M_{air}} \int_0^s \sigma_{i-air}\rho(s)ds \end{cases} \quad (F.22)$$

$v_s$ is the final root-mean-squared velocity, which is the initial status of volatiles at the start of diffusion, it was decelerated from $v_0$. Thus, the radius of deceleration zone $s$ of a specific kind of volatile molecules was able to calculate.

## 2.7 Fluid Field

The reference frame transfer need to be done before applying the diffusion model in a fluid field like geostrophic wind [20]. To fit the diffusion model in a pressure constant plane with hydrostatic equation (F.23), the vertical diffusion formulation (F.15) was written as (F.24), the horizontal diffusion formulation (F.16) was written as (F.25). The derivation was given in support information.

$$g\rho \nabla_P z = \nabla_z P \quad (F.23)$$

where $g$ is the gravitational acceleration, $\rho$ is the density of air, $\nabla_P z$ is the gradient of altitude $z$ on a pressure constant plane, $\nabla_z P$ is the gradient of pressure $P$ on a horizontal plane.

$$\left.\frac{\partial c_i}{\partial t}\right|_P^v = j_1 c_i + j_2 \frac{\partial c_i}{\partial z} + D_i \frac{\partial^2 c_i}{\partial z^2} \quad (F.24)$$

$$\left.\frac{\partial c_i}{\partial t}\right|_P^h = (\nabla_P z)^2 \left(k_1 c_i^3 + k_2 c_i^2 + k_3 c_i^1 + k_4 + k_5 c_i^{-1}\right) \quad (F.25)$$

where $j_1$ and $j_2$, $k_1$ to $k_5$ are coefficients listed at below.

$$j_1 = \frac{\partial D_i}{\partial T} \cdot \left(\frac{\partial T}{\partial z}\right)^2 - \frac{g\rho}{T} \cdot \frac{\partial D_i}{\partial P} \cdot \frac{\partial T}{\partial z} + \frac{D_i}{T} \cdot \frac{\partial^2 T}{\partial z^2}$$

$$j_2 = \left(\frac{\partial D_i}{\partial T} + 2D_i\right)\frac{\partial T}{\partial z} - g\rho \frac{\partial D_i}{\partial P}$$

$$k_1 = -k_2 = \frac{\partial D_i}{\partial P} \cdot \frac{M_i^2 g^2}{T(R - M_i)}$$

$$k_3 = \frac{D_i M_i^2 g^2 (M_i + 1)}{RT^2 (R - M_i)^2}$$

$$k_4 = -\frac{D_i M_i^2 g^2 (M_i + 2)}{RT^2 (R - M_i)^2}$$

$$k_5 = \frac{D_i M_i^2 g^2}{RT^2 (R - M_i)^2}$$

Thus, the concentration of specific volatile molecules at one point which is on a pressure constant plane can be calculated by integrating the above differential equation (F.24) and (F.25) with border and initial conditions.

Consider the diffusion of one point is in a quasi-geostrophic wind field [21], which is a moving reference frame. There are following components of motion $u_g$ and $v_g$ (F.26) (F.27).

$$u_g = -\frac{g}{2\Omega_p \sin\varphi} \cdot \frac{\partial Z}{\partial y} \qquad (F.26)$$

$$v_g = \frac{g}{2\Omega_p \sin\varphi} \cdot \frac{\partial Z}{\partial x} \qquad (F.27)$$

where $\Omega_p$ is the angular velocity vector of the planet, $\varphi$ is the latitude of the point. $Z$ is the height of the pressure constant plane. Thus, the concentration of volatiles at the point $(\Delta x, \Delta y)$ at time $t + \Delta t$, it should be calculated by applying the differential equation (F.24) (F.25) on point $(x', y')$, which

$$x' = \Delta x - \int_t^{t+\Delta t} u_g dt$$

$$y' = \Delta y - \int_t^{t+\Delta t} v_g dt$$

**2.8 Numerical Method**

The differential equation of concentration is a function of both pressure and temperature, which is a baroclinic system. The motion need to be described on different atmospheric levels. There is the horizontal momentum equation (F.28) at a point [22].

$$\frac{\partial}{\partial t}\left(-\frac{\partial \Phi_g}{\partial P}\right) + V \cdot \nabla\left(-\frac{\partial \Phi_g}{\partial P}\right) - \frac{B\omega}{g\rho^2} = \frac{1}{c_p \rho}\frac{dS}{dt} \qquad (F.28)$$

where $\Phi_g$ is geopotential, $V$ is the horizontal wind, $\nabla$ refers to differentiation at constant pressure, $B$ is the static stability parameter, which was represented by formulation (F.29). $\omega$ is the vector component vertical to ground surface, $\rho$ is the density of air, $c_p$ is the velocity of sound at pressure $P$, $S$ is the entropy.

$$B = -g\rho \frac{\partial \ln \theta}{\partial P} \qquad (F.29)$$

where $\theta$ is the potential temperature. Thus the full model in a fluid field was built.

## 3 Test and Result

**3.1 Baroclinic System Setting**

Limited to computational performance of currently used computer, a simple numerical temperature and pressure field on vertical axis was defined in this article for test, as it was shown in Table.2 [23]. Fig.4, Fig.5 and Fig.6 show the height correlated relation of temperature, pressure and density of air separately. The temperature and pressure were considered as constants in horizontal plane (F.30).

$$\nabla_z T = \nabla_z P = 0 \qquad (F.30)$$

Table.2 The "U.S. Standard Atmosphere 1976" is an atmospheric model of how the pressure, temperature, density, and viscosity of the Earth's atmosphere changes with altitude.

| H(m)  | T(°C)  | g(m/s$^2$) | P(10$^4$Pa) | ρ(kg/m$^3$) |
|-------|--------|-----------|-------------|-------------|
| 0     | 15.00  | 9.807     | 10.13       | 1.225       |
| 1000  | 8.50   | 9.804     | 8.988       | 1.112       |
| 2000  | 2.00   | 9.801     | 7.950       | 1.007       |
| 3000  | -4.49  | 9.797     | 7.012       | 0.9093      |
| 4000  | -10.98 | 9.794     | 6.166       | 0.8194      |
| 5000  | -17.47 | 9.791     | 5.405       | 0.7364      |
| 6000  | -23.96 | 9.788     | 4.722       | 0.6601      |
| 7000  | -30.45 | 9.785     | 4.111       | 0.5900      |
| 8000  | -36.94 | 9.782     | 3.565       | 0.5258      |
| 9000  | -43.42 | 9.779     | 3.080       | 0.4671      |
| 10000 | -49.90 | 9.776     | 2.650       | 0.4135      |
| 15000 | -56.50 | 9.761     | 1.211       | 0.1948      |
| 20000 | -56.50 | 9.745     | 0.5529      | 0.08891     |
| 25000 | -51.60 | 9.730     | 0.2549      | 0.04008     |
| 30000 | -46.64 | 9.715     | 0.1197      | 0.01841     |
| 40000 | -22.80 | 9.684     | 0.0287      | 0.003996    |
| 50000 | -2.5   | 9.654     | 0.007978    | 0.001027    |
| 60000 | -26.13 | 9.624     | 0.002196    | 0.0003097   |
| 70000 | -53.57 | 9.594     | 0.00052     | 0.00008283  |
| 80000 | -74.51 | 9.564     | 0.00011     | 0.00001846  |

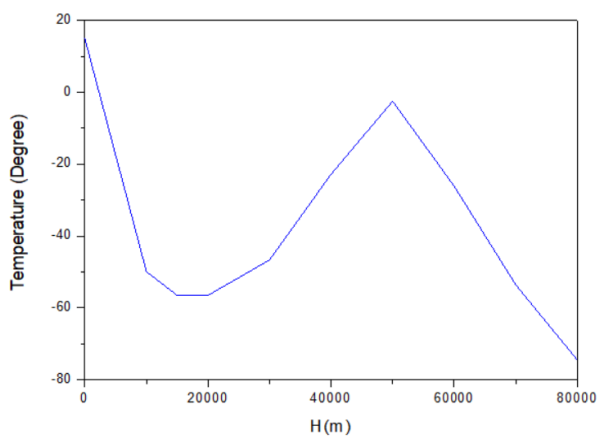

Fig.4 Temperature-height curve

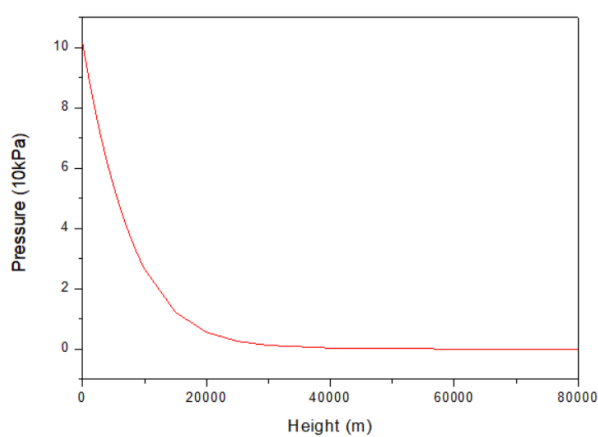

Fig.5 Absolute pressure-height curve

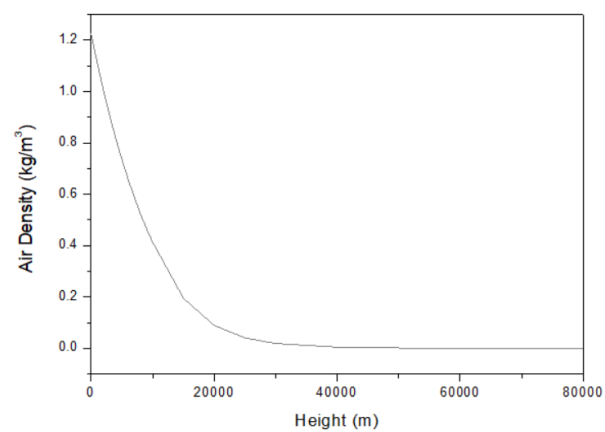

Fig.6 Air density-height curve

## 3.2 Chemical Composition Setting

The molar mass $M_i$ of specific volatiles molecules generated, mass abundance $A_M$ [11] and collision diameter d of molecules in this article were listed in Table.3, Table.4, Table.5, Table.6, Table.7, Table.8 and Table.9. The collision diameters were calculated by MOLCALC according to the structures of molecules given by citation [23].

Table.3 Components Group 1

|  | $P_4S_{10}$ | $P_4S_3$ | PN | $P_3N_5$ | $P_2O_5$ | $P_2O_3$ |
|---|---|---|---|---|---|---|
| $M_i$(g/mol) | 444 | 220 | 45 | 163 | 284 | 110 |
| $A_M(10^{-4})$ | 42 | 266 | 54 | 196 | 344 | 132 |
| $d(10^{-3}nm)$ | 615 | 332 | 184 | 552 | 520 | 335 |

Table.4 Components Group 2

|  | $PH_3$ | $PCl_3$ | $PCl_5$ | $S_2$ | $CS_2$ | $SO_2$ |
|---|---|---|---|---|---|---|
| $M_i$(g/mol) | 34 | 137 | 208 | 32 | 76 | 64 |
| $A_M(10^{-4})$ | 540 | 164 | 250 | 388 | 172 | 410 |
| $d(10^{-3}nm)$ | 208 | 366 | 410 | 134 | 360 | 221 |

Table.5 Components Group 3

|  | $SO_3$ | $H_2S$ | $SCl_2$ | $N_2$ | $(CN)_2$ | NO |
|---|---|---|---|---|---|---|
| $M_i$(g/mol) | 80 | 34 | 103 | 28 | 52 | 30 |
| $A_M(10^{-4})$ | 530 | 330 | 162 | 736 | 320 | 182 |
| $d(10^{-3}nm)$ | 303 | 134 | 203 | 146 | 387 | 119 |

Table.6 Components Group 4

|  | $NO_2$ | $H_3N$ | $N_2H_4$ | $NCl_3$ | CO | $CO_2$ |
|---|---|---|---|---|---|---|
| $M_i$(g/mol) | 46 | 17 | 32 | 120 | 28 | 44 |
| $A_M(10^{-4})$ | 282 | 194 | 104 | 160 | 112 | 180 |
| $d(10^{-3}nm)$ | 285 | 208 | 286 | 346 | 122 | 235 |

Table.7 Components Group 5

|  | $CH_4$ | $CCl_4$ | $C_2H_4$ | $C_3H_6$ | $C_5H_{12}$ | $C_2H_6$ |
|---|---|---|---|---|---|---|
| $M_i$(g/mol) | 16 | 119 | 28 | 42 | 16 | 30 |
| $A_M(10^{-4})$ | 482 | 64 | 112 | 172 | 64 | 236 |
| $d(10^{-3}nm)$ | 218 | 354 | 308 | 374 | 727 | 202 |

Table.8 Components Group 6

|  | $C_3H_8$ | $C_4H_{10}$ | $C_2H_2$ | $C_3H_4$ | $O_2$ | $O_3$ |
|---|---|---|---|---|---|---|
| $M_i$(g/mol) | 44 | 58 | 26 | 40 | 32 | 48 |
| $A_M(10^{-4})$ | 180 | 172 | 162 | 106 | 210 | 350 |
| $d(10^{-3}nm)$ | 417 | 544 | 333 | 424 | 144 | 238 |

Table.9 Components Group 7

|  | $H_2O$ | $ClO_2$ | $Cl_2O$ | $H_2$ | HCl | $Cl_2$ |
|---|---|---|---|---|---|---|
| $M_i$(g/mol) | 18 | 67 | 87 | 2 | 37 | 71 |
| $A_M(10^{-4})$ | 480 | 260 | 210 | 80 | 140 | 270 |
| $d(10^{-3}nm)$ | 153 | 248 | 276 | 71 | 132 | 202 |

The components of air in this article were set as 80% $N_2$ and 20% $O_2$, the collision diameter of air molecules was averaged base on concentration, which is 145.6 ($10^{-3}$ nm).

### 3.3 Vertical Concentration Distribution

By applying the data above on time of T1, the height correlated concentration distribution of each kind of volatile molecules was shown in Fig.7 to Fig.13, which were expressed by the ratio (0.05 to 1) of concentration at height h to initial concentration at h=0. The distribution was shown by gradient color from purple to orange.

T1 is the time of diffusion taken by the lightest component, $H_2$, which makes its concentration ratio reach 0.05 at 80000m. The data in Table.3 based on the atmosphere of Earth, the atmosphere can be divided by height, there is troposphere (0-11km), stratosphere (11-51km), mesosphere (51-71km) and ionosphere (above 71km) [24].

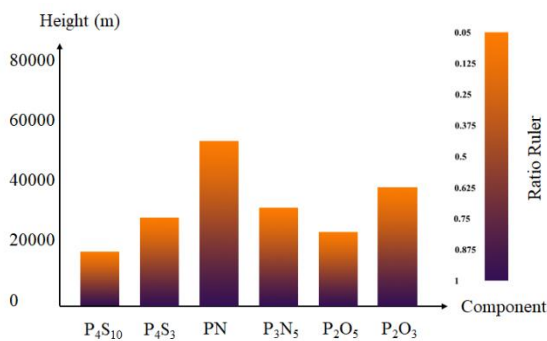

Fig.7 Concentration distribution of components Group 1

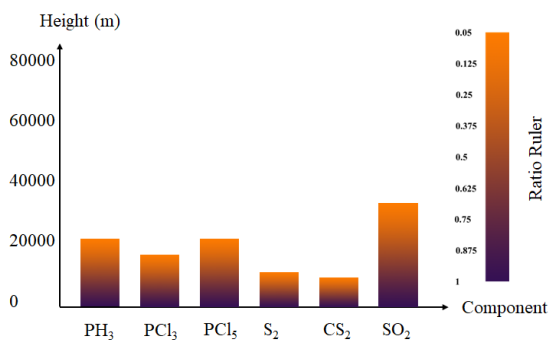

Fig.8 Concentration distribution of components Group 2

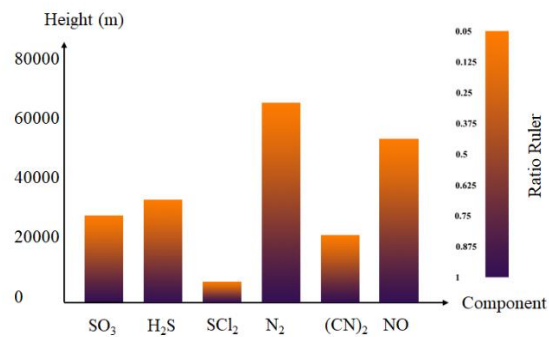

Fig.9 Concentration distribution of components Group 3

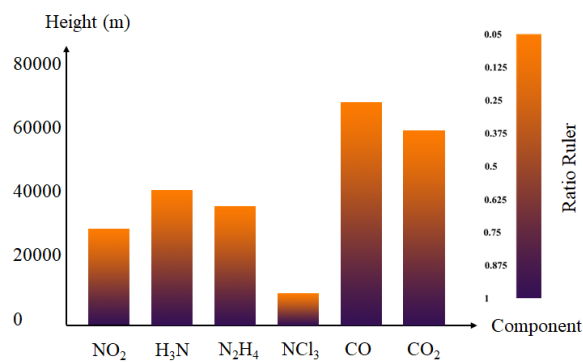

Fig.10 Concentration distribution of components Group 4

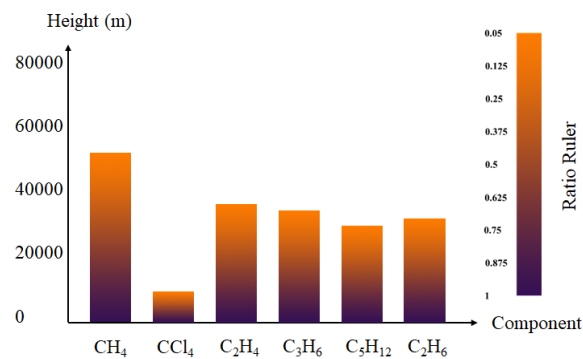

Fig.11 Concentration distribution of components Group 5

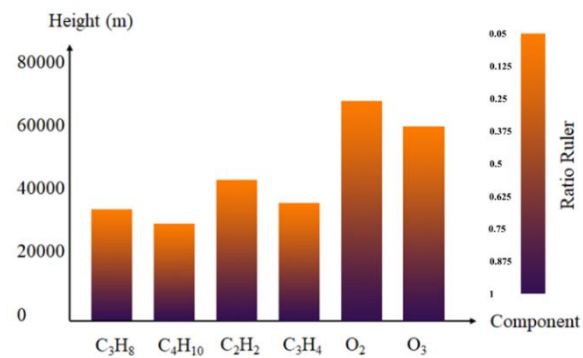

Fig.12 Concentration distribution of components Group 6

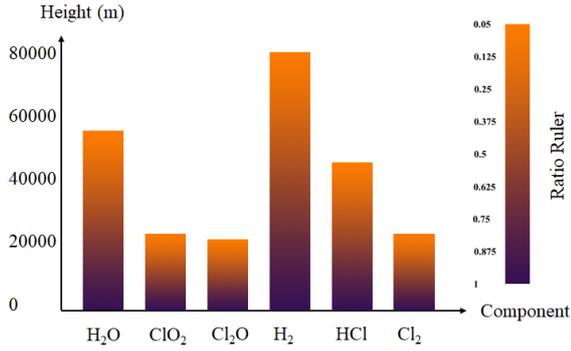

Fig.13 Concentration distribution of components Group 7

### 3.4 Horizontal Abundance Distribution

As the performance of calculation was limited, the horizontal distribution here in a fluid field was only considered under 10000m. A simple linear wind field was set (F.31), the wind speed $v_w$ varies with height h on x axis and keeps zero on y axis. The angular velocity of the planet was set as $7.292 \times 10^{-5}$ rad/s [25].

$$\vec{v_w} = \frac{2}{3000} h\vec{i} + 0\vec{j} \qquad (F.31)$$

By applying the data on time of T2, which is the time of diffusion taken by hydrogen, makes its abundance ratio reach 0.05 at 10000m. A $5500 \times 5500$ square meters horizontal plane was considered to be the diffusion area for calculation, it was divided by small areas with 1000m length to run model conveniently.

From Fig.4, Fig.5 and Fig.6, it can be seen that the temperature, pressure and air density are nearly linear function of height in troposphere under 10000m, which were expressed by lapse rate of temperature $\alpha$ (F.32), lapse rate of pressure $\beta$ (F.33), lapse rate of average air density $\gamma$ as formulation (F.34).

$$\frac{dT}{dz} = \alpha = -6.490 \times 10^{-3} \ °C/m \qquad (F.32)$$

$$\frac{dP}{dz} = \beta = -8.206 \ Pa/m \qquad (F.33)$$

$$\frac{d\rho}{dz} = \gamma = -8.115 \times 10^{-5} \ kg/m^4 \qquad (F.34)$$

The height correlated abundance distribution of each element was shown in Fig.14 to Fig.17, which were expressed by the ratio (0.05 to 1) of abundance at height h to initial abundance at h=0. The distribution was shown by gradient color from purple to orange. However, the computer we use currently is only able to deal the data along x and y axis, the data of the full $5500 \times 5500$ m$^2$ area will be done on higher performance computer in future work.

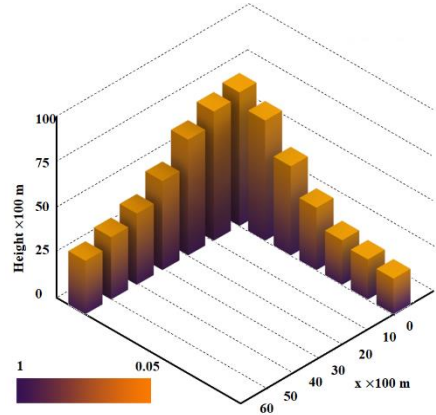

Fig.14 Abundance Distribution of P on x y axis

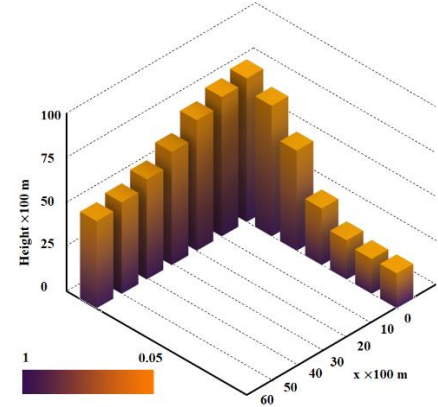

Fig.15 Abundance Distribution of S on x y axis

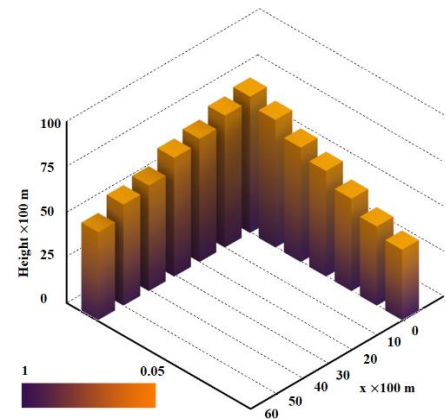

Fig.16 Abundance Distribution of N on x y axis

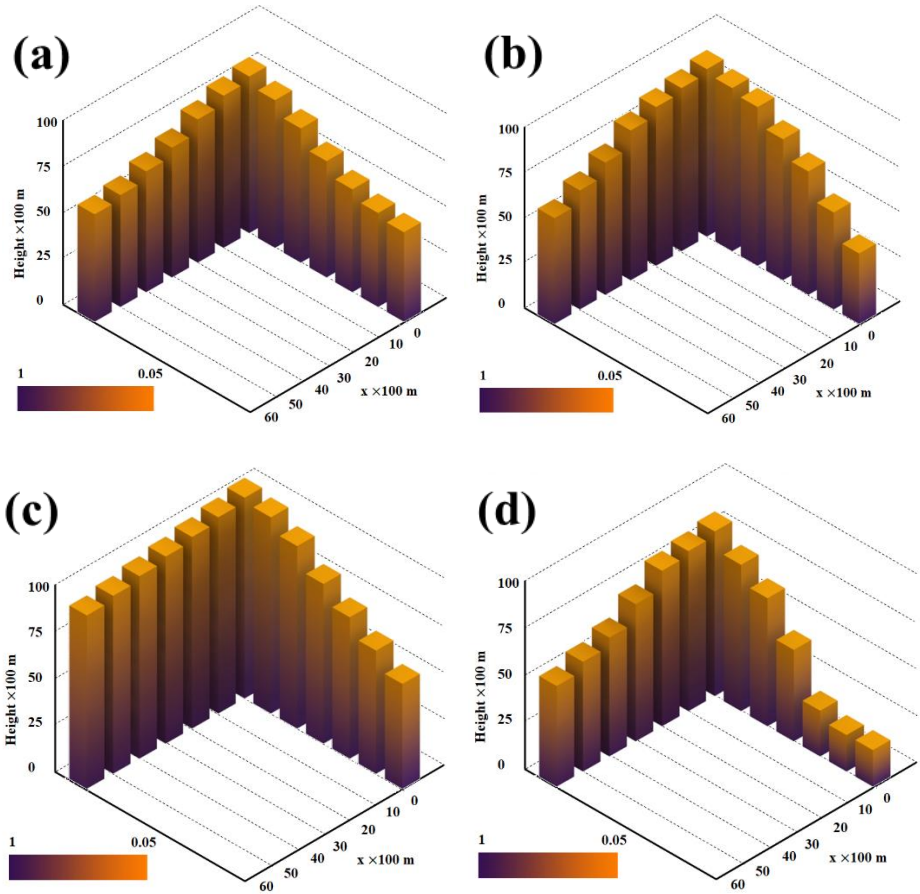

Fig.17 Abundance Distribution of C (a), O (b), H(c), Cl (d) on x y axis

### 3.5 Discussion

According to Fig.14 to Fig.17, it is obvious that most volatiles generated by volcano can keep large exist between 1000 and 5000 meters height in a 5500×5500 m² area when the lightest $H_2$ gas reaches 10000m height. There are elements tend to form light molar mass components, like H and O, which keep higher abundance in 5000 to 10000 meters height, the components they formed with less molar mass and smaller collision area make them easier to reach a higher concentration at a far horizontal range even at 5500m from the volatiles source. The height which matches median ratio of abundance was written as $R_{0.5}$. To hydrogen element in Fig.17(c), $R_{0.5}(y=0m)$ is 5041m, $R_{0.5}(y=5500m)$ is 2618m, which decreases 48.07% during a 5500m horizontal diffusion range. To oxygen element in Fig.17(b), $R_{0.5}(y=0m)$ is 4349m, $R_{0.5}(y=5500m)$ is 1824m, which decreases 58.06% during a 5500m horizontal diffusion range. On the other hand, there is obvious abundance decreasing trend on the y axis of P, S and Cl in Fig.14, Fig.15 and Fig.17(d), which is the diffusion without motion field except rotation of the planet. The reason is that they are tend to form higher molar mass and larger collision area components, which caused the diffusion coefficient $D_i$ decrease to keep them only diffuse in a shorter range during the time we set. To sulfur element, $R_{0.5}(y=0m)$ is 3861m, $R_{0.5}(y=5500m)$ is 909m, which decreases 76.46% during a 5500m horizontal diffusion range. To chlorine element, which is the heaviest element considered in this article, $R_{0.5}(y=0m)$ is 4091m, $R_{0.5}(y=5500m)$ is only 817m, which decreases 80.03% during a 5500m horizontal diffusion range.

However, the fluid field (x axis) might cause huge change which could promote the horizontal diffusion of volatiles. To chlorine element in Fig.17(d), $R_{0.5}(x=0m)$ is 4091m, $R_{0.5}(x=5500m)$ is 2678m, which decreases 34.53% during a 5500m horizontal diffusion range with the wind field we set, this is far from the previous decreasing percentage 80.03% on y axis without the fluid field, which means that the fluid field increases the difference 45.50% during a 5500m horizontal diffusion of chlorine element. To hydrogen element in Fig.17(c), $R_{0.5}(x=0m)$ is 5041m, $R_{0.5}(x=5500m)$ is 4309m, which decreases 14.52% during a 5500m horizontal diffusion range with the wind field. The previous decreasing percentage on y axis of hydrogen is 48.07%, the difference between them is 33.55%. To compare hydrogen with chlorine, heavier molar mass components or elements were influenced by fluid field stronger in diffusion.

Table.10 $R_{0.5}$ on x, y axis of elements

| meters | P | S | C | N | O | H | Cl |
|---|---|---|---|---|---|---|---|
| $R_{0.5}(x_0,y_0)$ | 3418 | 3861 | 3977 | 3528 | 4349 | 5041 | 4091 |
| $R_{0.5}(x_{5500})$ | 1273 | 2368 | 2547 | 2273 | 2742 | 4309 | 2678 |
| $R_{0.5}(y_{5500})$ | 918 | 909 | 2252 | 1811 | 1824 | 2618 | 817 |

Thus, the fluid field motion like convection can strongly promote the diffusion of heavy molar mass volatiles components to bring them into troposphere elements cycle, instead of depositing near the volcano vent and the rapid attenuation of abundance along horizontal distance. Fluid field makes concentration of heavy volatiles dilute horizontally and harder to keep exist over 50000m. According to Fig.9, Fig.13, the light molar mass volatiles like $N_2$, $H_2$ can keep higher relative exist over 15000m by self-diffusion, which makes them are the main components generated by volcanism and finally take participate in stratosphere, mesosphere and ionosphere cycle.

## 4 Conclusion

In this article, a simplified four processes model was built to evaluate the diffusion and contribution to atmosphere elements abundance of volatiles components which were generated by volcanism. The modeling results on a single vertical axis with 80000m height and a 5500×5500 m$^2$ area with 10000m height showed that the vertical abundance distribution highly depends on the coefficients of diffusion of molecules that the elements formed. The diffusion coefficients are mainly influenced by molar mass and average collision diameter of molecules. The horizontal abundance distribution depends on the temperature and pressure of atmosphere more likely.

However, the model was not able to extend on a larger area due to the low performance of the computer used currently. The example fluid field applied in this article is also too far from a realistic motion field in atmosphere. Although this article is not able to give a numerical prediction of the contribution of volcanism volatiles to atmospheric elements abundance distribution on a specific planet, it still prove that the four processes model built and talked in this article is an practicable idea and able to be tested on data which based on observation.

## Reference

[1] E. S. K. e. al (2009). "GEODYNAMICS AND RATE OF VOLCANISM ON MASSIVE EARTH-LIKE PLANETS." The Astrophysical Journal 700.
[2] Kerber, L., et al. (2009). "Explosive volcanic eruptions on Mercury: Eruption conditions, magma volatile content, and implications for interior volatile abundances." Earth and Planetary Science Letters 285(3-4): 263-271.
[3] Zimbelman, J. R., et al. (1991). "THE EVOLUTION OF VOLCANISM, TECTONICS, AND VOLATILES ON MARS - AN OVERVIEW OF RECENT PROGRESS." Proceedings of Lunar and Planetary Science 21: 613-626.
[4] Turbet, M., et al. (2018). "Modeling climate diversity, tidal dynamics and the fate of volatiles on TRAPPIST-1 planets." Astronomy & Astrophysics 612.
[5] Schindler, T. L. and J. F. Kasting (2000). "Synthetic spectra of simulated terrestrial atmospheres containing possible biomarker gases." Icarus 145(1): 262-271.
[6] Zolotov, M. Y. (2011). "On the chemistry of mantle and magmatic volatiles on Mercury." Icarus 212(1): 24-41.
[7] J.W., W. L. H. I. (1981). Journal of Geophysical Research 86(B4): 30.
[8] Wilson, L. and K. Keil (1997). "The fate of pyroclasts produced in explosive eruptions on the asteroid 4 Vesta." Meteoritics & Planetary Science 32(6): 813-823.

[9] Head, J. W., et al. (2008). "Volcanism on Mercury: Evidence from the first MESSENGER flyby." Science 321(5885): 69-72.

[10] Pint, B. A., et al. (1998). An oxygen potential gradient as a possible diffusion driving force. Materials-Research-Society Symposium on Diffusion Mechanisms in Crystalline Materials at the MRS Spring Meeting, San Francisco, Ca.

[11] ZH., et al. (2022). "Thermal dynamics modeling on composition of volatiles generated by volcanism"

[12] Anikin, Y. A., et al. (2015). "Method of calculating the collision integral and solution of the Boltzmann kinetic equation for simple gases, gas mixtures and gases with rotational degrees of freedom." International Journal of Computer Mathematics 92(9): 1775-1789.

[13] Diaz, C. and R. A. Olsen (2009). "A note on the vibrational efficacy in molecule-surface reactions." Journal of Chemical Physics 130(9).

[14] An, F. H., et al. (2022). "Effect of stress, concentration and temperature on gas diffusion coefficient of coal measured through a direct method and its model application." Fuel 312.

[15] Paradisi, P., et al. (2001). "A generalized Fick's law to describe non-local transport effects." Physics and Chemistry of the Earth Part B-Hydrology Oceans and Atmosphere 26(4): 275-279.

[16] Cussler, E. L. (1997). Diffusion: Mass Transfer in Fluid Systems (2nd ed.). New York: Cambridge University Press. ISBN 0-521-45078-0.

[17] Welty, James R.; Wicks, Charles E.; Wilson, Robert E.; Rorrer, Gregory (2001). Fundamentals of Momentum, Heat, and Mass Transfer. Wiley. ISBN 978-0-470-12868-8.

[18] Hirschfelder, J.; Curtiss, C. F.; Bird, R. B. (1954). Molecular Theory of Gases and Liquids. New York: Wiley. ISBN 0-471-40065-3.

[19] Munoz, A. G. (2007). "Formulation of molecular diffusion in planetary atmospheres." Planetary and Space Science 55(10): 1414-1425.

[20] Roth, R., et al. (1999). "Geostrophic wind, gradient wind, thermal wind and the vertical wind profile - A sample analysis within a planetary boundary layer over Arctic sea-ice." Boundary-Layer Meteorology 92(2): 327-339.

[21] Shen, J., et al. (1999). "On a wind-driven, double-gyre, quasi-geostrophic ocean model: Numerical simulations and structural analysis." Journal of Computational Physics 155(2): 387-409.

[22] Grubisic, V. and M. W. Moncrieff (2000). "Parameterization of convective momentum transport in highly baroclinic conditions." Journal of the Atmospheric Sciences 57(18): 3035-3049.

[23] Engineering ToolBox, (2003). U.S. Standard Atmosphere vs. Altitude. [online] Available at: https://www.engineeringtoolbox.com/standard-atmosphere-d_604.html [Accessed 05 Apr. 2022].

[24] Lam, M. M., et al. (2014). "Solar wind-driven geopotential height anomalies originate in the Antarctic lower troposphere." Geophysical Research Letters 41(18): 6509-6514.

[25] Brumberg, V. A. and E. Groten (2001). "A note on Earth's rotation angular velocity in the general-relativity framework." Journal of Geodesy 75(12): 673-676.